\documentclass[twoside,a4paper,12pt]{article}

\usepackage[english]{babel}
\usepackage{amsmath}
\usepackage{amssymb}
\usepackage{amsthm}
\usepackage{mathrsfs}
\usepackage{pdfsync}

\newcommand{\pd}[2]{\frac{\partial {#1}}{\partial {#2}}}
\newcommand{\be}{\begin{equation}}
\newcommand{\ee}{\end{equation}}
\newcommand{\n}{\noindent}
\newcommand{\RE}{\mathbb R}
\newcommand{\CO}{\mathbb C}
\newcommand{\ulim}{\operatorname{u}-\lim}
\newcommand{\sgn}{\operatorname{sgn}\,}

\newcommand{\ve}{\varepsilon}

\newcommand{\ga}{\gamma}
\newcommand{\la}{\lambda}

\newcommand{\al}{\alpha}

\renewcommand{\Im}{\operatorname{Im}\,}

\newcommand{\GG}{\mathcal G}
\newcommand{\oOmega}{\overline \Omega}
\newcommand{\oV}{\overline V}
\newcommand{\Hboh}{H^{\ga}}
\newcommand{\Rboh}{R^{\ga,\ve}}
\newcommand{\oH}{\overline H}
\newcommand{\oR}{\overline R}
\newcommand{\tM}{\widetilde M}
\newcommand{\tB}{\widetilde B}
\newcommand{\tm}{\widetilde m}
\newcommand{\tvarphi}{\tilde\varphi}
\newcommand{\hla}{\hat\lambda}
\newcommand{\OO}{\mathcal{O}}

\newtheorem{theorem}{Theorem}[section]
\newtheorem{lemma}{Lemma}[section]
\newtheorem{proposition}{Proposition}
\theoremstyle{definition}

\author{Claudio Cacciapuoti$^{1}$ and Pavel Exner$^{2,3}$}
\title{Nontrivial edge coupling from a Dirichlet network squeezing:
the case of a bent waveguide}
\date{\small $^1$Institut f\"ur Angewandte Mathematik, Universit\"at Bonn,
Wegelerstr. 6, 53115 Bonn, Germany \\
$^{2}$Nuclear Physics Institute, Czech Academy of Sciences,
25068 \v{R}e\v{z} near Prague \\
$^{3}$Doppler Institute, Czech Technical University,
B\v{r}ehov\'{a} 7, 11519 Prague, Czechia \\
\emph{caccia@na.infn.it, exner@ujf.cas.cz}}

\begin{document}

\maketitle

\begin{abstract}
\noindent In distinction to the Neumann case the squeezing limit
of a Dirichlet network leads in the threshold region generically
to a quantum graph with disconnected edges, exceptions may come
from threshold resonances. Our main point in this paper is to show
that modifying locally the geometry we can achieve in the limit a
nontrivial coupling between the edges including, in particular,
the class of $\delta$-type boundary conditions. We work out an
illustration of this claim in the simplest case when a bent
waveguide is squeezed.
\end{abstract}

\section{Introduction}

Quantum mechanics on graphs attracted a lot of attention recently
-- let us just mention the proceeding volume \cite{BCFK} as a
guide to the abundant bibliography in the field. The interest has
different sources, important among them are numerous existing and
potential applications. While simple and versatile, however,
quantum graph models have a problem: in a sense they offer too
much freedom. The requirement of self-adjointness, or probability
current conservation, determines a class of boundary conditions
which connect the column vectors $\Psi$ and $\Psi'$ of boundary
values of the wave functions and their derivatives at a given
graph vertex. Following \cite{Ha, KS} these conditions can be cast
into the unique form
 \begin{equation}\label{param}
 (U-I)\Psi+i(U+I)\Psi'=0\,,
 \end{equation}
where $U$ is an $n\times n$ unitary matrix, $n$ being the number
of edges sprouting of the vertex.

Hence a vertex coupling contains $n^2$ free parameters to be
fixed. Asking about the meaning of various couplings one can
naturally get useful insights by obtaining boundary conditions
through limit of families of (regular or singular) interactions on
the graph \cite{Ex, CE, ET}. A proper understanding of the problem
requires, however, to find an interpretation of the coupling in
terms of models without free parameters; a natural idea is to
investigate the motion of a free quantum particle on a system of
thin tubes which shrink towards the graph.

This is a longstanding and nontrivial problem and the answer
depends substantially on the network dynamics we work with. In the
case when the Hamiltonian (which can be identified with the
Laplacian by a suitable choice of units) refers to tubes with
Neumann boundary, the limit yields typically quantum graphs with
free boundary conditions (often called not quite appropriately
Kirchhoff) in the vertices, described by $U= \frac2n\, \mathcal{J}
-I$, where $\mathcal{J}$ is the $n\times n$ matrix whose all
entries are equal to one -- see \cite{FW, KZ, RS, Sa}, the same is
true also for shrinking families of ``sleeve'' manifolds without a
boundary \cite{EP05}; the convergence is norm-resolvent
\cite{Po05a} and the conclusion extends to resonances on such
structures \cite{EP07}.

From the viewpoint of application to semiconductor structures and
similar objects, however, it is the case of Dirichlet (hard-wall)
boundary which is more important. It is very different from its
Neumann counterpart and more difficult and relevant results
started appearing only recently. The main source of the
difficulties is that in the Dirichlet case geometric perturbations
like bending, ``swelling'', twisting or branching give rise to an
effective interaction -- see, e.g., \cite{ES, DE, EEK} and
references therein -- sometimes attractive, sometimes repulsive,
which changes spectral and scattering properties of such networks.
Even the statement of the problem is more complicated than in the
Neumann case where we naturally investigate spectrum around the
zero value which is the continuum threshold. The analogous
quantity in a Dirichlet network blows up to infinity and one has
to choose the reference point by a suitable energy
renormalization. In most cases, with the notable exception of the
recent paper \cite{MV}, the attention is paid to the vicinity of
the ``running'' threshold.

It is generally conjectured, that the generic limit in the
vicinity of the threshold corresponds to a fully decoupled graph
with Dirichlet conditions at the edge endpoints. This is obvious
if the vertex regions squeeze faster than tubes as in \cite{Po05b}
but it is expected to hold even without such an additional
effective repulsion. To see the reason one has to realize that the
problem at hand can be by scaling rephrased as analysis of a network
of tubes of constant cross section whose overall size grow. This
means that the distances between the ``vertices'' tend to infinity
and the character of solutions to the Schr\"odinger equation is
determined by its asymptotic properties around each single vertex.
Away of the threshold, as in the particular case of \cite{MV}
cited above, the limiting boundary conditions are generally
nontrivial and given by the scattering properties of the ``fat star''
region. Around the threshold, on the other hand, the scattering is
generically suppressed in view of the mentioned effective
interaction -- see a brief discussion in \cite{MV}, the
forthcoming paper \cite{Gr} and also a related problem with
Dirichlet boundary replaced by a confining potential in \cite{DT}
-- leading to the Dirichlet decoupling.

This limiting behaviour is not universal, though. The situation
changes if the described system associated with the vertex
possesses a threshold resonance. The simplest case where a
nontrivial effect of this type can be observed is a bent tube
which squeezes in the limit to a graph of two halfline edges
joined in a single vertex \cite{ACF}: in presence of a threshold
resonance one gets the line with a point interaction of
scale-invariant type -- cf.~\cite{HC} and references therein. More
general results of that type were announced in \cite{Gr}.

Our main point in this paper is to argue than one can go further
and construct classes of squeezing Dirichlet networks which
produce wider families of vertex coupling when renormalized to the
continuum threshold, including those with nonempty discrete
spectrum or resonances. The procedure we propose consists of two
steps:

\begin{description}

\item{(i)} choose a network collapsing to a graph in such a way
that the limit Hamiltonian has a threshold
resonance\footnote{The original network Hamiltonian at that may or
may not have such a resonance, depending on the limiting procedure
used. A threshold resonance may be present, e.g., in the leading
term of the perturbation expansion w.r.t. the squeezing parameter
as in the model discussed below.}

\item{(ii)} change the scaling properties of the vertex region
slightly, typically by adding higher order terms in the scaling
parameter

\end{description}

The modification in point (ii) can be achieved in various ways,
for instance, one can ``wiggle'' the edges angles or scale the
vertex volume region at a rate which differs from that of the
``edge tubes'' by a higher order term, a combination of such
perturbations, etc. Incidentally, the same effect can also be
obtained by introducing suitable potentials into the vertex
region, but a purely geometric way is probably the most
interesting.

Notice that the described approximation follows the same scheme
which one uses when interpreting pseudopotentials, or point
interactions in dimensions two and three, by a suitable nonlinear
scaling starting from a threshold resonance \cite{AGH-KHII}. 
On the other hand, there is a
large difference between the two cases coming from the fact that
the geometric approximation discussed here covers not a single
operator class but a broad variety of systems. Consequently, the
proposal made above has a status of \emph{a conjecture} and the
corresponding procedure must be made concrete and worked out
properly in each particular case.

To show that this programme is not void we are going in the rest
of the paper perform this task for the bent-waveguide system
studied in \cite{ACF}. We will show that modifying the bending
angle around a threshold resonance value we can arrive at the
limit at a two-parameter class of point interactions on the line
including the important particular case of the $\delta$
interaction. In our model the system has multiple
threshold resonances and the limiting procedure described above can
be associated with each of them, we expect that this is a standard behavior of networks with Dirichlet boundary. 
We will describe the model and state the results in
the next section. Then we extend the analysis of short-range
potentials from \cite{ACF} to more general scaling, and in the
last section we will prove our main theorem.

\setcounter{equation}{0}
\section{The bent-waveguide model and the results}

\n As we have said we use the model studied in \cite{ACF}, namely
a planar waveguide of constant width obtained by ``fattening'' a
fixed smooth curve along its normal. We denote by $C$ a curve
embedded in $\RE^2$, i.e. $C:=\{(x,y)\in\RE^2|
\,x=\ga_1(s),\,y=\ga_2(s),\,s\in\RE\}$, assuming that it is
parameterized by its arc length, $\ga_1'^2+\ga_2'^2=1$. Moreover,
we denote by $\ga(s)$ the signed curvature of $C$,
\[
\ga(s):=\ga'_{2}(s)\ga''_{1}(s) -\ga'_{1}(s)\ga''_{2}(s)\,;
\]
it completely characterizes the curve $C$ up to Euclidean
transformations and the curvature radius at a given point is
given by $r=|\ga|^{-1}$. We suppose that the curve is not
self-intersecting, i.e., it has no loops, and for simplicity we
consider only curves with a compactly supported curvature. The
last assumption means that the curve $C$ is made up of two
straight half lines joined by a smooth curve. In particular, the
(overall) bending angle of $C$ is the angle between the two half
lines, which is equal to $\theta=\int_\RE\ga(s)\,ds\,.$

The above mentioned fat curve which is our waveguide is the open
set $\Omega\in\RE^2$ defined as
\[
\Omega:=\{(x,y)\in\RE^2|\,x= \ga_1(s)-u\ga_2'(s),
y=\ga_2(s)+u\ga_1'(s), s\in\RE,u\in(-d,d)\}\,,
\]
where $s$ and $u$ represent a global system of coordinates in
strip, $s$ being the coordinate along the curve while $u$ is the
distance along the normal to $C$. The width of the waveguide is
constant and equal to $2d$ where $d>0$. Another standard
assumption we made is that $d$ is smaller than the curvature
radius, $d\|\ga\|_\infty<1$. The closure of $\Omega$ is
conventionally denoted by $\oOmega$. The (negative) Laplacian with
Dirichlet boundary conditions on $\partial\oOmega$, denoted as
$-\Delta_\Omega^D$, is the Friedrichs extension of the positive,
symmetric operator $L_0:=-\Delta$ with $\mathscr
D(L_0):=C_0^\infty(\Omega)$.

The main geometric object of our study will be a family of
waveguides whose shape and width depend on a scaling parameter
$0<\ve\leqslant1$ according to
 \be \label{scaling}
\ga_\ve(s):=\frac{\sqrt{\la(\ve)}}{\ve}\ga\Big(\frac{s}{\ve}\Big)\,;\qquad
d_\ve:=\ve^\al d\,,\quad\textrm{with }\al>1\,, \ee
where $\la(\ve)$ is a fixed function to be specified below; its
presence is the main difference comparing to \cite{ACF} because of
the term $\sqrt{\la(\ve)}$. We suppose that $\la(\ve)$ is real,
positive and analytic near the origin, and moreover, that it
expands as
 \be
\label{hpla}
\la(\ve)=1+\la_1\ve+\OO(\ve^2)\qquad\textrm{with}\quad\la_1=\la'(0)\,.
\ee
We assume that the condition $d_\ve \|\ga_\ve\|_\infty <1$ is satisfied
for every $0<\ve\leqslant\ve_0$ with some
$\ve_0>0$. The scaling \eqref{scaling} gives rise to a family of
curves, $C_\ve:=\{(x,y)\in\RE^2|\,x =\ga_{\ve,1}(s),\,y
=\ga_{\ve,2}(s),\,s\in\RE\}$, the bending angle $\theta_\ve$ of
which changes slightly with respect to $\ve$,
\[
\theta_\ve=\int_\RE \ga_\ve(s)\,ds
=\theta\sqrt{\la(\ve)}=\theta\bigg(1
+\frac12{\la_1}\ve\bigg)+\OO(\ve^2)\,.
\]
They in turn generate a family of bent waveguides, i.e. domains
$\Omega_\ve$ defined by
\[
\Omega_\ve:=\{(x,y)\in\RE^2|\,x
= \ga_{\ve,1}(s)-u\ga_{\ve,2}'(s),y
=\ga_{\ve,2}(s)+u\ga_{\ve,1}'(s), s\in\RE,u\in(-d_\ve,d_\ve)\}\,.
\]
In the limit $\ve\to0$ the strip family shrinks to a graph,
denoted by $\GG$, made up of two  edges and one vertex. Our aim is
to investigate the limit of the respective operator family
$-\Delta_{\Omega_\ve}^D$ when $\ve\to0$. We will show that it
approximates in a suitable sense an operator on $\GG$, namely the
Schr\"odinger operator on the line with a point interaction depending
on $\ga$ and $\la(\ve)$.

Before to state our main theorem, let us introduce some notation
and mention some preliminary facts. Writing for brevity
$\Omega'=\RE\times(-d ,d )$, we recall the following result
\cite{DE, ES}:
\begin{proposition}\label{prop1}
For any $0<\ve\leqslant\ve_0$ let $C_\ve$ be as described above,
with $\ga$ piecewise $C^2$ and compactly supported, such that
$\ga',\ga''$ are bounded. Then $-\Delta^D_{\Omega_\ve}$ is
unitarily equivalent to the operator $H_\ve$ defined as the
closure of the e.s.a. operator $H_{0\ve}$ acting on $L^2(\Omega')$
as
\[
H_{0\ve}=-
\pd{}{s}\frac{1}{(1+\ve^{\al-1} u\sqrt{\la(\ve)}\gamma(s/\ve))^2}\pd{}{s}-
 \frac{1}{\ve^{2\al}}\pd{^2}{u^2}+\frac{1}{\ve^2} V_\ve(s,u)\,,
\]
with the effective potential
\[
V_\ve(s,u)=-\frac{\la(\ve)\gamma(s/\ve)^2}
{4(1+\ve^{\al-1}u\sqrt{\la(\ve)}\gamma(s/\ve))^2}
+\frac{\ve^{\al-1} u\sqrt{\la(\ve)}\gamma''(s/\ve)}
{2(1+\ve^{\al-1}u\sqrt{\la(\ve)}\gamma(s/\ve))^3}
-\frac{5}{4}\frac{\ve^{2\al -2} u^2\la(\ve)\gamma'(s/\ve)^2}
{(1+\ve^{\al-1}u\sqrt{\la(\ve)}\gamma(s/\ve))^4}
\]
and ${\mathscr D}(H_{0,\ve})=\{ \psi\in L^2(\Omega')|\, \psi\in
C^\infty(\Omega') \,,\,\psi(s,d)=\psi(s,-d)=0 \, , \, H_{0\ve}
\psi \in L^2(\Omega') \}$.
\end{proposition}

\n Let us next introduce the transversal modes, i.e., the
normalized functions $\phi_n(u)$ which solve the equation
$-{\ve}^{-2\al}\phi''_n(u)=E_{\ve,n}\phi_n(u)$ with the boundary
conditions $\phi_n(\ve^\al d)=\phi_n(-\ve^\al d)=0$. In
particular, the corresponding eigenvalues $E_{\ve,n}$ are
explicitly given by
\[
E_{\ve,n}=\Big(\frac{n\pi}{2d\ve^\al}\Big)^2\qquad
\textrm{with}\quad n=1,2,\dots\,.
\]
The resolvent of $H_\ve$ admits an integral representation with
the kernel $( H_{\ve} - z)^{-1} (s,u,s',u')$ for every
$z\in\rho(H_\ve)$ with $\Im \sqrt{z}>0$, where $\rho(H_\ve)$ is
the resolvent set of $H_\ve$, cf.~Thm~II.37 in \cite{Si}. Using it
we define the projection of the resolvent on the normal modes
eigenspaces as
\[
\oR_{n,m}^{\ve}(k^2, s,s') :=  \int_{-d}^{d}du\,du' \,
\phi_{n}(u)  ( H_{\ve} - k^2 - E_{\ve,m} )^{-1}(s,u,s',u')
\phi_{m} (u')\,.
\]
The operators $\oR_{n,m}^{\ve}(k^2) : L^2(\RE) \rightarrow
\textrm{Ran}\big[\,\oR_{n,m}^{\ve}(k^2)\big]\subset L^2(\RE)$
introduced in this way are bounded operator-valued analytic
functions of $k^2$ for all $k^2\in\CO\backslash\RE$ and $\Im k>0$.

Next we have to recall some facts about one-dimensional
Schr\"odinger operators. We say that the Hamiltonian
\be \label{oH} \oH=-\frac{d^2}{ds^2}+\oV(s) \ee
has a zero energy resonance if there exist a function $\psi_r\in
L^\infty(\RE)$, $\psi_r\notin L^2(\RE)$, such that $\oH\psi_r=0$
holds in the sense of distributions. In particular, if
\be \label{hpoV} \int_\RE\oV(s)\,ds\neq0\quad\textrm{and}\quad
e^{a|\cdot|}\oV\in L^1(\RE) \ee
holds for some $a>0$, then exactly one of the following situations
can occur \cite{BGW}:

\begin{description}

\item\emph{Case I}: The Hamiltonian $\oH$ does not have a zero
energy resonance.

\item\emph{Case II}: The Hamiltonian $\oH$ has a zero energy
resonance; in such a case the function $\psi_r$ can be chosen real
and two real constants can be defined,
\be \label{c1c2} c_1=\bigg[\int_\RE\oV(s)ds\bigg]^{-1}
\int_\RE\int_\RE\oV(s)\frac{|s-s'|}{2}\oV(s')\psi_r(s')\,ds\,ds'\,,
\qquad c_2=-\frac{1}{2}\int_\RE s\,\oV(s)\psi_r(s)\,ds\,, \ee
and moreover, $c_1$ and $c_2$ cannot vanish simultaneously. Let us
stress that the constants $c_1$ and $c_2$ defined in \eqref{c1c2}
coincide with those employed in \cite{ACF}.

\end{description}

Let us next introduce a pair of Hamiltonians on $\GG$ both acting
as $f\mapsto -f''$ but differing by the boundary conditions in the
vertex. The first is the Dirichlet-decoupled operator $\oH^d$ with
the domain $\mathscr D(\oH^d) := \{ f\in H^2(\RE \setminus 0 )
\cap H^1(\RE)|\,f(0)=0 \}$. The other is a point-interaction
Hamiltonian $\oH^r$, which again acts as $\overline{H}^{r} f=
-f''$ but on the domain
\[
\begin{aligned}
&\mathscr D(\oH^r) := \bigg\{ f\in H^2(\RE \setminus 0 )| \,
 (c_1 + c_2 ) f(0^+ ) = (c_1 - c_2 ) f(0^- )\, ,\\
&\qquad\qquad\qquad(c_1 - c_2 ) f' (0^+ ) = (c_1 + c_2 ) f' (0^- )
+\frac{\hla}{c_1+c_2}f(0^-) \bigg\}\quad \textrm{for}\;c_2\neq-c_1 \,;\\
&{\mathscr D}(\oH^r):= \bigg\{f\in H^2(\RE \setminus 0 )|\,
 f(0^-) = 0
\,, \, f' (0^+ ) = \frac{\hla}{4c_1^2}f(0^+)
\bigg\}\quad \textrm{for}\;c_2=-c_1\,,
\end{aligned}
\]
where we put
\be \label{hla} \hla:=\la_1\int_\RE \oV(s)\big(\psi_r(s)\big)^2\,ds\,. \ee
The graph $\GG$ identifies naturally with a line and both $\oH^d$
and $\oH^r$ belong to the family of self-adjoint extensions of the
symmetric operator $\overline L_0f:=-f''$ with the domain
$\mathscr D(\overline L_0):=C_0^\infty(\RE\backslash\{0\})\:$
\cite{ABD}.

Let us say a few more words on the family $\oH^r$ which obviously
depends on two real parameters. It is a straightforward exercise
to check that the boundary conditions appearing in the definition
of $\mathscr D(\oH^r)$ can be rewritten in the form (\ref{param})
with $\Psi:=(f(0^+),f(0^-))^T$, $\Psi':=(f'(0^+),-f'(0^-))^T$ and the $2\times 2$ unitary matrix
 \be \label{u} U:= \frac{1}{2(c_1^2+c_2^2)+ i\hla}
 \left( \begin{array}{cc}
 -4c_1c_2 -i\hla & 2(c_1^2-c_2^2) \\
 2(c_1^2-c_2^2) & 4c_1c_2 -i\hla \end{array} \right)\,. \ee
In particular, for $\la_1=0$ the boundary conditions define the
``scale invariant'' Hamiltonian obtained in \cite{ACF}.
Applications of this point interaction were discussed recently in
\cite{HC}, and it is worth mentioning that it appears also in the
theory of regular tree graphs \cite{So}. On the other hand, in
distinction to \cite{ACF} we have here a wider class which
contains, in particular, the standard $\delta$ interaction of
coupling strength $\hla\:$ \cite{AGH-KHII} corresponding to
$c_1=1$ and $c_2=0$. Spectral and scattering properties of $\oH^r$ are
well known \cite{EG} and we recall them only briefly:

 \begin{proposition}
For any $-\infty<\hla\leqslant\infty$, the essential spectrum of
$\oH^r$ is absolutely continuous and coincides with the interval
$[0,\infty)$. Furthermore, for $\hla>0$  there are no eigenvalues,
while for $\hla<0$ there is just one negative eigenvalue given by
$k^2=k_0^2=-\frac14 \hla^2(c_1^2+c_2^2)^{-1}$ and the
corresponding normalized eigenfunction is
\[
\psi_0=\sqrt{\frac{|\hla|}{2}}\frac{1}{c_1^2+c_2^2}\left\{\begin{aligned}
&(c_1-c_2)e^{ik_0s}&\quad s>0\\
&(c_1+c_2)e^{-ik_0s}&\quad s<0
\end{aligned}\right\}\,, \qquad k_0=\frac{i|\hla|}{2(c_1^2+c_2^2)}\,,
\;\:\hla<0\,.
\]
Finally, for $\hla=0$ the operator $\oH^r$ has a zero energy
resonance. The on-shell scattering matrix at energy $k^2,\:
k\geqslant0$ is given by $\mathscr{S}(k)=
\begin{bmatrix}
\mathscr{T}^l(k)&\mathscr{R}^r(k)\\
\mathscr{R}^l(k)&\mathscr{T}^r(k)
\end{bmatrix}$ with the amplitudes
 \[
 \mathscr{T}^{\{l,r\}}(k)=\frac{2k(c_1^2-c_2^2)}{2k(c_1^2+c_2^2)+i\hla}\,,
 \qquad \mathscr{R}^{\{l,r\}}(k)=
 \pm \frac{4kc_1c_2 \mp i\hla}{2k(c_1^2+c_2^2)+i\hla}\,.
 \]
\end{proposition}

Let now $G_k$ be the resolvent of the free Laplacian on $\RE$, it is a bounded operator-valued analytical function of $k^2$
for $k^2\in \CO\backslash\RE^+$ and $\Im k>0$, with the integral
kernel given by
\[
G_k(s-s') = \frac{i}{2k} e^{ik|s-s'|}\qquad
k^2\in\CO\backslash\RE^+,\;\Im k>0\,.
\]
By Krein's formula \cite{ABD, EG} the integral kernel of the
resolvent $\oR^{d}(k^2):=(\oH^d-k^2)^{-1}$ is
\[
\oR^{d}(k^2,s,s')=G_k(s-s')+2ikG_k(s)G_k(s')\;,\quad
k^2\in\CO\backslash\RE^+\,,\;\Im k>0\,,
\]
while the integral kernel of the resolvent
$\oR^{r}(k^2):=(\oH^r-k^2)^{-1}$ equals
\[
\begin{aligned}
\overline R^r(k^2;s,s')=&G_k(s-s')+
2ik\frac{2kc_2^2+i\hla}{2k(c_1^2+c_2^2)+i\hla}G_k(s)G_k(s')
+\frac{4ic_2^2}{2k(c_1^2+c_2^2)+i\hla}G_k'(s)G_k'(s')\\
&+\frac{4kc_1c_2}{2k(c_1^2+c_2^2)+i\hla}\big[G_k(s)G_k'(s')
+G_k'(s)G_k(s')\big]\;,\quad k^2\in\rho(\oH^r)\,,\;\Im k>0\,.
\end{aligned}
\]
Our main result can be now stated as follows:
\begin{theorem}\label{main}
Suppose that for every $0<\ve\leqslant\ve_0$ the curve $C_\ve$ has
no self-intersections, $\ga $ is piecewise $C^2$ with a compact
support, and $\ga',\ga''$ are bounded. Assuming $\al > 5/2$, we have:

\begin{description}

\item (i) If $\displaystyle -\frac{d^2}{ds^2}-\frac14 {\ga^2(s)}$
does not have a zero energy resonance, then
\[
\ulim_{\ve \rightarrow 0} \oR_{n,m}^{\ve}(k^2) =
\delta_{n,m}\oR^d(k^2) \qquad k^2\in\CO\backslash\RE,\;\:\Im
k>0\,.
\]
\item (ii) If, on the other hand, $\displaystyle
-\frac{d^2}{ds^2}-\frac14 {\ga^2(s)}$ has a zero energy resonance,
then
\[
\ulim_{\ve \rightarrow 0} \oR_{n,m}^{\ve}(k^2) = \delta_{n,m}\oR^r(k^2)
 \qquad k^2\in\CO\backslash\RE,\;\:\Im k>0\,,
\]
where the constants $c_1$, $c_2$ and $\hla$, defined in
\eqref{c1c2} and \eqref{hla}, are obtained by setting
$\oV=-\frac14 \ga^2$ and $\delta_{n,m}$ indicates the Kronecker
symbol, $\delta_{n,m}=0$ if $n\neq m$ and $\delta_{n,n}=1$.

\end{description}

\end{theorem}

\setcounter{equation}{0}
\section{The limit of short range potentials in dimension one}

The main ingredient in the proof of Theorem~\ref{main} is the
analysis of scaling properties of one dimensional Hamiltonians.
Specifically, we will find the limiting behaviour as $\ve\to 0$
for
\[
\oH_\ve: =-\frac{d^2}{ds^2}+
\frac{\la(\ve)}{\ve^2}\oV\Big(\frac{s}{\ve}\Big)\,,\qquad s\in\RE\,.
\]
Recall that this problem is well studied if the limit is
considered, roughly speaking, around the free operator
\cite{AGH-KHII}. The case which involves threshold resonances is
different and in a sense similar to approximations of point
interactions in dimension two and three mentioned above. It is
useful to discuss this issue separately because in our opinion it
is of independent interest as approximation of a class of point
perturbations of the Laplacian in dimension one with scaled
potentials. Let us stress that the $\delta'$-type interactions do
not belong to this class -- a way to approximate them by regular
potentials can be found in \cite{ENZ}. The main idea of our
analysis comes from the work \cite{BGW}.

In the following we suppose that the conditions \eqref{hpoV} are
satisfied. As $\la(\ve)$ is real analytic near the origin by
assumption we can make the expansion (\ref{hpla}) for small $\ve$
more specific writing
 \be \label{hpla2} \la(\ve)=1+\sum_{n=1}^{\infty}\la_n\ve^n\,. \ee
For every $\ve>0$ the resolvent of $\oH_\ve$ is a bounded
operator-valued analytical function of $k^2$ as long as
$k^2\in\CO\backslash\RE^+$, $k^2\notin\sigma_p(\oH_\ve)$ and $\Im
k>0$, where $\sigma_p(\oH_\ve)$ denotes the point spectrum of
$\oH_\ve$. As usual we factorize the interaction using the
functions
\[
v(s):=|\oV(s)|^{1/2}\,,\quad u(s):=\sgn[\oV(s)]|\oV(s)|^{1/2}\,,
\]
which allows us to write $(\oH_\ve-k^2)^{-1}$ as in
\cite{AGH-KHII}, namely
 \be \label{swinging} (\oH_\ve-k^2)^{-1}=
G_k-\frac{\la(\ve)}{\ve}A_\ve(k)T_\ve(k)C_\ve(k)\,, \ee
where
\[
T_\ve(k)=\big[1+ \la(\ve)u G_{\ve k} v\big]^{-1} \qquad \Im k
\geqslant 0, \;k\neq 0, \;k^2\notin \sigma_p(\overline{H_\ve} )
\]
and  $A_\ve(k),\: C_\ve(k)$ are defined via their integral
kernels, $A_\ve(k;s,s')=G_k(s-\ve s')v(s')$ and
$C_\ve(k;s,s')=u(s)G_k(\ve s-s')$, respectively. We are interested
in the behaviour of $T_\ve(k)$ as $\ve\to0$. To this aim we define
the operators $P$ and $Q$ by
\[
P:=\frac{1}{(v,u)}(v,\cdot\,)u\;,\quad Q:=1-P
\]
where $(\cdot\,,\cdot)$ denotes the standard scalar product in
$L^2(\RE)$; let us notice that by assumption we have
$(v,u)=\int_\RE\oV(s)ds\neq0$. The operator $T_\ve(k)$ can be
written as in \cite{BGW},
 \be
 \label{stravagante} T_\ve(k)= \bigg[1+ \frac{i(v,u)}{2\ve k}P
 +\tM_\ve(k)\bigg]^{-1} \ee
where $\tM_\ve(k)\in\mathscr{B}(L^2,L^2)$, the Banach space of
bounded operators from $L^2(\RE)$ to $L^2(\RE)$, for every $\ve>0$
and $\Im k>0$. Furthermore, if $e^{a|\cdot|}\overline V\in
L^1(\RE)$ holds for some $a>0$ then $\tM_\ve$  is analytic with
respect to $\ve$ for $\ve>-a/(2\Im k)$ and the following series
expansion converges in the $\mathscr{B}(L^2,L^2)$-norm,
\[
\tM_\ve(k)=\sum_{n=0}^\infty\ve^{n}\tm_n(k)\,,
\]
\n where
\[
\tm_n(k):=(ik)^nm_n+\frac{i\la_{n+1}(v,u)}{2k}P
+\sum_{j=0}^n\la_{n-j}(ik)^jm_j\qquad n=0,1,2,\dots\,.
\]
The operators $m_n$ are Hilbert-Schmidt and do not depend on $k$,
their integral kernels being given by the expressions
\[
m_n(s,s') = -u(s)\, \frac{|s-s'|^{n+1} }{2(n+1)!}\, v(s')\,.
\]
The behaviour of $T_\ve(k)$ as $\ve\to0$ depends strongly  on the
presence of a zero energy resonance for the Hamiltonian $\oH$.
Under the assumptions \eqref{hpoV} the presence of such a
resonance is equivalent to the existence of a function $\varphi_0
\in L^2(\RE)$ which satisfies the relation
\be \label{eqphi0} \varphi_0 + Q M_0 Q \varphi_0=0\,. \ee
Furthermore, if such a $\varphi_0$ exists, it can  be chosen real,
in which case  the constants $c_1$, $c_2$ and $\hla$ defined in
\eqref{c1c2} and \eqref{hla}, respectively, are related to
$\varphi_0$ by
\[
c_1=\frac{(v,m_0\varphi_0)}{(v,u)}\;,\quad
c_2=\frac{1}{2}((\cdot)v,\varphi_0)\;,\quad
\hla=\la_1(\tvarphi_0,\varphi_0)\,.
\]
with $\tvarphi_0(s):=\sgn[\oV(s)]\varphi_0$, and $u(s)\psi_r(s)
=-\varphi_0(s)$ holds a.e. -- cf.~Lemma ~2.2. in \cite{BGW}. Let
us introduce the operator
\[
P_0:=\left\{
\begin{aligned}
&0\qquad&&\textrm{in the case I}\\
&\frac{(\tvarphi_0,\cdot)\varphi_0}{(\tvarphi_0,\varphi_0)}&&\textrm{in the case II}
\end{aligned}
\right.
\]
and the complementary projection $Q_0:=1-P_0$. From Lemma~3.1 in
\cite{BGW} we infer that for $\ve\in\CO\backslash\{0\}$ small
enough the following norm convergent series expansion holds,
\[
[1+Qm_0Q+\ve]^{-1}
=\frac{P_0}{\ve}+\sum_{n=0}^\infty(-\ve)^nT_{red}^{n+1}\,,
\]
where $T_{red}=\ulim_{\ve\to0}[1+Qm_0Q+\ve]^{-1}Q_0$ is the
reduced resolvent. The following claim is a generalization of
Theorem~3.1 in \cite{BGW}
\begin{lemma}
\label{newlemma} Suppose that $\oV$ satisfies the conditions
\eqref{hpoV} and take $\la(\ve)$  real analytic near the origin
and with the series expansion \eqref{hpla2}. Assume that
$k^2\notin\sigma_p(\oH_\ve)$, $\Im k>0$ and additionally, that in
the case II $k\neq-i\hla/(2(c_1^2+c_2^2))$. Then for all $\ve$
small enough the operator $T_\ve(k)$ has the following
norm-convergent series expansions
\be \label{luz} T_\ve(k)=\sum_{n=p}^\infty\ve^nt_n(k)\,, \ee
where $p=0$ in the case I and $p=1$ in the case II. Moreover, we
have

\begin{description}
\item (i) In the case I
\be \label{forest1}
(v,t_0u)=0\,;\;\big((\cdot)v,t_{0}u\big)=\big(v,t_{0}(\cdot)u\big)=0\,;
\ee
 \be \label{forest2} (v,t_1u)=-2ik\,. \ee
\item (ii) In the case II
  \be \label{free}
t_{-1}u=t_{-1}^*v=0\,;\;\big((\cdot)v,t_{-1}(\cdot)u\big)
=-\frac{4ic_2^2}{2k(c_1^2+c_2^2)+i\hla}\,; \ee
 \be \label{free2}
(v,t_0u)=0\,;\;\big((\cdot)v,t_{0}u\big)=\big(v,t_{0}(\cdot)u\big)
=\frac{4kc_1c_2}{2k(c_1^2+c_2^2)+i\hla}\,;
\ee
 \be \label{free3}
(v,t_1u)=-2ik\frac{2kc_2^2+i\hla}{2k(c_1^2+c_2^2)+i\hla}\,. \ee
\end{description}
\end{lemma}
\begin{proof}
We prove the lemma first in the case II. Let us assume that the
equation \eqref{eqphi0} is solved by $\varphi_0\in L^2(\RE)$. By
using the relation \cite{BGW}
\[
1+\frac{i(v,u)}{2\ve k}P=\bigg[Q+\frac{2\ve k}{2\ve k+i(v,u)}P\bigg]^{-1}
\]
in formula \eqref{stravagante} we obtain
 \be \label{secrets}
T_\ve=\bigg[1+Q\tM_\ve+\frac{2\ve k}{2\ve
k+i(v,u)}P\tM_\ve\bigg]^{-1}\bigg[Q+\frac{2\ve k}{2\ve
k+i(v,u)}P\bigg]\,. \ee
Since $Q\tm_0=Qm_0$ the following norm convergent series expansion
holds \cite{BGW}
 \be \label{africa}
\big(1+Q\tm_0+\delta\big)^{-1}=\bigg[\frac{P_0}{\delta}
+\sum_{n=0}^{\infty}(-\delta)^nT_{red}^{n+1}\bigg]\bigg
[1-\frac{Qm_0P}{1+\delta}\bigg]\,.
\ee
Taking into account that  $P_0-P_0Qm_0P=-P_0m_0$ and performing a
simple manipulation, we can set  $\delta=-2i\ve k/(v,u)$ and use
relation \eqref{africa} in formula \eqref{secrets} to obtain
\[
\begin{aligned}
T_\ve=&\bigg[1+\bigg(\frac{(v,u)}{2i\ve k}P_0m_0+D_\ve
\bigg)\bigg(\frac{2i\ve k}{(v,u)}+Q\tM_\ve^{(1)}
+\frac{2\ve k}{2\ve k+i(v,u)}P\tM_\ve\bigg)\bigg]^{-1}\\
&\times \bigg[\frac{(v,u)}{2i\ve k}P_0m_0+D_\ve\bigg]
\bigg[Q+\frac{2\ve k}{2\ve k+i(v,u)}P\bigg]\,,
\end{aligned}
\]
where
\[
D_\ve(k):=\frac{2\ve k}{2\ve k+i(v,u)}P_0Qm_0P+
\sum_{n=0}^\infty\bigg(\frac{2i\ve
k}{(v,u)}\bigg)^nT_{red}^{n+1}\bigg[1-\frac{Qm_0P}{1-2i\ve
k/(v,u)} \bigg]
\]
and $\displaystyle\tM_\ve^{(j)}(k) =\sum_{n=j}^\infty
\ve^{n}\tm_n(k)$ with $j=1,2,\dots\;$. After some computation we
arrive at the following formula for the operator $T_\ve$,
\be \label{joe}
T_\ve=\big[1+P_0\tB+E_\ve\big]^{-1}\bigg[\frac{(v,u)}{2i\ve
k}P_0m_0+D_\ve\bigg]\bigg[Q+\frac{2\ve k}{2\ve k+i(v,u)}P\bigg]\,,
\ee
where
\[
\tB(k)=m_0+\frac{(v,u)}{2ik}m_0Q\tm_1(k)-m_0P\tm_0(k)
\]
and
\[
\begin{aligned}
E_\ve(k)=&\frac{(v,u)}{2i\ve k}P_0m_0Q\tM_\ve^{(2)}(k)
+\frac{2\ve k}{2\ve k+i(v,u)}P_0m_0P\tm_0(k)
-\frac{i(v,u)}{2\ve k+i(v,u)}P_0m_0P\tM_\ve^{(1)}(k)+\\
&+
D_\ve(k)\bigg(\frac{2i\ve k}{(v,u)}+Q\tM_\ve^{(1)}(k)
+\frac{2\ve k}{2\ve k+i(v,u)}P\tM_\ve(k)\bigg)\,.
\end{aligned}
\]
The operator $\tB(k)$ does not depend on $\ve$ while $D_\ve(k)$
and $E_\ve(k)$ have with respect to the parameter the following
norm convergent series expansions
\be \label{three} D_\ve(k)=\sum_{n=0}^\infty\ve^nd_n(k)\;,\quad
E_\ve(k)=\sum_{n=1}^\infty\ve^ne_n(k)\,. \ee
Let us notice that
\[
\frac{(\tvarphi_0,\tB\varphi_0)}{(\tvarphi_0,\varphi_0)}
=-1-\frac{(v,u)}{(\tvarphi_0,\varphi_0)}
(c_1^2+c_2^2+i\hla/(2k))
\]
and $P_0\tB P_0=(\tvarphi_0,\tB\varphi_0)/ (\tvarphi_0,
\varphi_0)P_0$. In a similar way as in \cite{BGW} we can
explicitly evaluate for $k\neq-i\hla/(2(c_1^2+c_2^2))$ the inverse
of $1+P_0\tB$ obtaining
\[
[1+P_0\tB]^{-1}=1+\frac{(\tvarphi_0,\varphi_0)}{(v,u)}
\frac{1}{c_1^2+c_2^2+i\hla/(2k)}P_0\tB\,.
\]
Formula \eqref{joe} implies that in the case II the norm
convergent series expansion \eqref{luz} holds for $\ve$ small
enough with $p=-1$.

Keeping only the terms corresponding of order $\ve^{-1}$ at the
right hand side of equation \eqref{joe} and using the relation
$P_0m_0Q=-P_0$ we obtain
\[
t_{-1}=\frac{(v,u)}{2ik}[1+P_0\tB]^{-1}P_0m_0Q
=\frac{(\tvarphi_0,\varphi_0)}{2ik}
\frac{1}{(c_1^2+c_2^2+i\hla/(2k))}P_0\,.
\]
Relations \eqref{free} follow from $P_0u=P_0^*v=0$ and
$((\cdot)v,P_0(\cdot)u)=4c_2^2/(\tvarphi_0,\varphi_0)$. Inspecting
the terms of order zero in $\ve$ at the right hand side of
\eqref{joe} we obtain
\[
t_0=[1+P_0\tB]^{-1}
[-P_0m_0P+d_0Q-e_1t_{-1}]\,.
\]
The relation $(v,t_0u)=0$ is a consequence of the fact that
$\big([1+P_0\tB]^{-1}\big)^*v=v$. Moreover, by a direct
calculation based on the relation $T_{red}P=PT_{red}=P$ similar to
\cite{BGW} one can check that
\[
((\cdot)v,t_0u)=-\big((\cdot)v,[1+P_0\tB]^{-1}P_0m_0u\big)
\]
and
\[
(v,t_0(\cdot)u)=-(v,e_1t_{-1}(\cdot)u)
\]
from which  the relations \eqref{free2} follow. Formula
\eqref{free3} is obtained by considering the terms of order $\ve$
at the right hand side of equation \eqref{joe} in combination with
the relation
\[
(v,t_1u)=-\frac{2ik}{(v,u)}(v,d_0u)-(v,e_1[1+P_0\tB]^{-1}P_0m_0u)\,.
\]

It remains to deal with the case I, in such a case $P_0=0$ and the equation
\eqref{joe} becomes
 \be \label{joe2}
T_\ve=\big[1+E_\ve\big]^{-1}D_\ve\bigg[Q+\frac{2\ve k}{2\ve
k+i(v,u)}P\bigg]\,, \ee
where
\[
D_\ve(k)=
\sum_{n=0}^\infty\bigg(\frac{2i\ve k}{(v,u)}\bigg)^nT_{red}^{n+1}
\bigg[1-\frac{Qm_0P}{1-2i\ve k/(v,u)}
\bigg]
\]
and
\[
E_\ve(k)=D_\ve(k)\bigg(\frac{2i\ve k}{(v,u)}+Q\tM_\ve^{(1)}(k)
+\frac{2\ve k}{2\ve k+i(v,u)}P\tM_\ve(k)\bigg)\,.
\]
The series expansions \eqref{three} still hold, and the norm
convergent series expansion \eqref{luz} in case I is valid with
$p=0$. Let us notice that $[1+E_\ve]$ is invertible for
$\ve\geqslant0$ with $\ve$ small enough, and consequently, it is
not necessary to assume $k\neq-i\hla/(2(c_1^2+c_2^2))$. In the
case I we thus have
\[
t_0=d_0Q=T_{red}[1-Qm_0P]Q\,,
\]
from which it easily follows that $(v,t_0u)=0$, and from
$PT_{red}=T_{red}P=P$ one gets relations \eqref{forest1}. The
terms of order $\ve$ at the right hand side of equation
\eqref{joe2} give
\[
t_1=\frac{2k}{i(v,u)}d_0P+d_1Q-e_1d_0Q\,,
\]
the formula \eqref{forest2} then  follows from $(v,d_0u)=(v,u)$.
\end{proof}

With this result at hand we can follow the argument line of
\cite{ACF} to establish the norm resolvent convergence of the
Hamiltonian $\oH_\ve$ to $\oH^d$ or $\oH^r$, depending on the
potential $\oV$; we omit the details. Using formulae
\eqref{forest1}--\eqref{free3} in the proof of Lemma~1 of
\cite{ACF} we arrive at the following conclusion:
\begin{theorem}\label{1d}
Suppose that $\oV$ satisfies the conditions \eqref{hpoV} and
$\la(\ve)$ is real analytic near the origin having the series
expansion \eqref{hpla2}. Then we have

\begin{description}

\item (i) In the case I
\[
\ulim_{\ve\to0} (\overline H_\ve-k^2)^{-1}=\oR^d(k^2) \qquad
k^2\in\CO\backslash\RE,\;\Im k>0\,.
\]
\item (ii) In the case II
\[
\ulim_{\ve\to0} (\overline H_\ve-k^2)^{-1}=\oR^r(k^2) \qquad
k^2\in\CO\backslash\RE,\;\Im k>0\,.
\]
\end{description}
\end{theorem}

\setcounter{equation}{0}
\section{Proof of Theorem \ref{main}}

Also the rest of the proof of the main result now follows closely
\cite{ACF} so it is sufficient to sketch the argument. It splits
into two steps. The first was dealt with in the previous section,
the second step consists of the proof of the claim given below.
Since the latter is essentially as Lemma 3 in \cite{ACF}, we just
state it omitting the details.
\begin{lemma}\label{lemma3}
Suppose that $C_\ve$ has no self-intersections for every
$0<\ve\leqslant\ve_0$, and moreover, that $\ga $ is piecewise
$C^2$, has compact support. and $\ga',\ga''$ are bounded. Fix an
$\al > 5/2$ and define $\Hboh_{\ve}$ as the closure of the e.s.a.
operator
\[
\Hboh_{0\ve}:= - \pd{^2}{s^2}- \frac{1}{\ve^{2\al}}  \pd{^2}{u^2}
-\frac{\la(\ve)}{\ve^2} \frac{\ga(s/\ve)^2}{4}
\]
with the domain
\[
{\mathscr D}(\Hboh_{0\ve}):=\{\psi\in L^2(\Omega' )|\,
\psi\in C^\infty(\Omega' )\,,\,
\psi(s,d)=\psi(s,-d)=0 \, , \,
\Hboh_{0\ve} \psi \in L^2(\Omega') \}\,.
\]
Defining the matrix elements $\Rboh_{n,m} $ with respect to the
transverse modes $\phi_{n}$ and  $\phi_{m}$,
\[
\Rboh_{n,m} (k^2;s,s') =  \int_{-d}^{d}du\,du' \phi_{n}(u)  (
\Hboh_{\ve} - k^2 - E_{\ve,m} )^{-1}(s,u,s',u') \phi_{m} (u')\,,
\]
we have
\[
\ulim_{\ve \rightarrow 0}  \big(R_{n,m}^{\ve}(k^2) -\Rboh_{n,m}
(k^2)\big)  = 0
\qquad k^2\in\CO\backslash\RE,\;\Im k>0\,.
\]
\end{lemma}

\n In analogy with \cite{ACF} Theorem \ref{main} is now obtained by
combination of Theorem \ref{1d} and Lemma \ref{lemma3}.

\bigskip \noindent \emph{Acknowledgment:} The research was partially
supported by GAAS and MEYS of the Czech Republic under projects
A100480501 and LC06002, by the Collaborative Research Center
(SFB) 611 ``Singular Phenomena and Scaling in Mathematical Models'' and by the Deutsche Akademische Austauschdienst (DAAD).

\end{document}